\documentclass[aps,prd,a4paper,superscriptaddress,nofootinbib,
showkeys,amsfonts,amssymb,amsmath,11pt]{revtex4}
\usepackage{amssymb,latexsym}
\usepackage{amsmath, amsthm}
\usepackage{amscd}
\usepackage{amsthm}
\usepackage{times}
\usepackage{epsfig}
\usepackage{psfrag}
\usepackage{graphicx}
\usepackage{amssymb,latexsym}

\begin{document}

\title{The Rainbows of Gravity}

\author{Pankaj S. Joshi} \footnote{14th Vaidya-Raychaudhuri
Endowment Award Lecture, given at the IAGRG Conference at
H. N. Bahuguna Garhwal University, Srinagar; March 2013}
\email{psj@tifr.res.in}

\affiliation{Tata Institute of Fundamental Research, Homi 
Bhabha road, Colaba, Mumbai 400005, India}

\def\m{$M$}
\def\mb{$\overline M$}
\def\i{$\cal I^+$}
\def\s{spacetime}
\def\p{singularity}
\def\t{\triangle}

\begin{abstract} We present here a spectrum of developments 
and predictions in gravitation theory in recent years which
appear to be amongst some of the most exciting directions. These 
include the spacetime singularities, gravitational collapse 
final states, and the deep cosmic conundrums that the new results 
on these issues have revealed. Amongst these are the cosmic 
censorship and the paradox of predictability in the universe, 
and the possible emerging implications for a quantum theory 
of gravity. The likely contact with observations and 
implications for relativistic astrophysics and black hole 
physics today are indicated. 

\end{abstract}
\maketitle

\section{Introduction}
After Einstein proposed the general theory of 
relativity describing the gravitational force in terms of spacetime 
curvatures, the proposed field equations related the spacetime 
geometry to the matter content of the universe. 
In general relativity, the universe is modeled as a spacetime, 
which has mathematically a structure of a four dimensional 
differentiable manifold. This means that locally the spacetime 
is always flat, in a sufficiently small region around any point, 
but on a larger scale it does not have to be so and it can have 
more rich and varied structure. A two-dimensional example of 
such a manifold is a sphere, which is flat enough in the vicinity 
of any single point on its surface, but has a non-zero global 
curvature.

The earliest solutions found for the field equations 
were the Schwarzschild metric 
representing the gravitational field around an isolated body 
such as a spherically symmetric star, and the Friedmann cosmological 
models. Both these contained a spacetime singularity where 
the curvatures and energy densities were 
infinite and the physical description would then break down. 
In the Schwarzschild solution such a singularity was present 
at the center of symmetry $r=0$ whereas for the Friedmann 
models it is found at the epoch $t=0$ which is beginning 
of the universe and origin of time where the scale factor 
for the universe vanishes and all objects are crushed to a 
zero volume due to infinite gravitational tidal forces.

Even though the physical problem posed by the existence 
of such a strong curvature singularity was realized immediately 
in these solutions, which turned out to have several important 
implications towards the experimental verification of the
general relativity theory, initially this phenomenon was 
not taken seriously. It was generally thought that the existence 
of such a singularity must be a consequence of the very 
high degree of symmetry imposed on the 
spacetime while deriving these solutions. Subsequently, 
the distinction between a genuine spacetime singularity and 
a mere coordinate singularity became
clear and it was realized that the singularity at $r=2m$ 
in the Schwarzschild spacetime was only a coordinate singularity 
which could be removed by a suitable coordinate transformation. 
It was clear, however, that the genuine curvature singularity at
$r=0$ cannot be removed by any such transformations. 
The  hope was then that when more general solutions are considered 
with a less degree of symmetry requirements, such singularities 
will be avoided. This issue was sorted out when a detailed study of 
the structure of a general spacetime and the associated 
problem of singularities was taken up by Hawking, Penrose, 
and Geroch (see for example, 
Hawking and Ellis, 1973),
which showed that singularities are in fact a much more
general phenomena in gravitation theories.

Further to the general relativity theory in 1915, the 
gravitation physics was a relatively quiet 
field with few developments till about 1950s. However, 
the 1960s saw the emergence of new observations in high energy 
astrophysics, such as quasars and high energy phenomena at 
the center of galaxies such as extremely energetic jets. 
These observations, together with important theoretical 
developments such as studying the global structure of spacetimes 
and singularities, led to important results in black 
hole physics and relativistic astrophysics and cosmology.

My purpose here is to indicate and highlight a spectrum 
of such developments and results which deal with probably 
some of the most exciting current issues on which useful 
research in gravitation and cosmology is centered today. 
This is of course a personal perspective and no claim to 
completeness is made. However, I hope that what is presented 
below will paint an interesting view of the landscape 
of gravity physics and the emerging cosmic frontiers. While 
doing so, I discuss what I think to be rather interesting 
results, including some of our work on the final endstates 
of gravitationally collapse, cosmic censorship, 
and black holes and naked singularities. Related major 
cosmic conundrums such as the issue of predictability in 
the universe are discussed, and observational implications 
of naked singularities are indicated.

\section{Spacetime singularities}
As mentioned above, the work in early 1970s in 
gravitation theories showed that a spacetime will admit
singularities within a rather general framework provided it 
satisfies certain reasonable physical assumptions such as 
the positivity of energy, a suitable causality assumption and 
a condition implying strong gravitational fields, 
such as the existence of trapped surfaces.
It thus followed that singularities form a rather general 
feature of the relativity theory. In fact, these considerations  
ensure the existence of singularities in other theories 
of gravity also which are based on a spacetime manifold 
framework and that satisfy the general conditions 
such as those stated above.

Therefore the scenario that emerges is, 
essentially for all classical spacetime theories of gravitation, 
the occurrence of singularities form an inevitable and integral
part of the description of the physical reality. In the vicinity
of such a singularity, typically the energy densities, spacetime 
curvatures, and all other physical quantities would blow up,
thus indicating the occurrence of super ultra-dense regions 
in the universe. The behaviour of such regions may not be 
governed by the classical theory itself, which may breakdown 
having predicted the existence of the singularities, and 
a quantum gravitational theory would be the likely 
description of the phenomena created by such singularities.

Firstly, it is to be clarified how to characterize
a spacetime singularity. It turns out that it is the 
notion of  geodesic incompleteness that characterizes 
a singularity in an effective manner for a \s\  and enables  
their existence to be proved by means of general 
enough theorems, which involve a consideration of the 
gravitational focusing caused by the \s\  curvature in 
congruences of timelike and null geodesics. This turns out 
to be the main cause of the existence of  \p\ in the form 
of non-spacelike incomplete geodesics in \s . 
The issue of physical nature of 
a \s\  \p\ is important.  There are many types of singular 
behaviours possible for a \s\  and some of these could be regarded 
as  mathematical pathologies in the  \s\  rather than having 
any physical significance. This will be especially so if the \s\  
curvature and similar other physical quantities remained finite 
along an incomplete non-spacelike geodesic
in the limit of approach to the \p. A singularity will be
physically important when there is a powerful enough 
curvature growth along  singular geodesics, and the physical 
interpretation and implications of the same are
to be considered.

Considering various situations, the occurrence of nonspacelike 
geodesic incompleteness has been generally agreed upon as the 
criterion for the existence of a \p\ for a \s .
It is clear that if a \s\  manifold contains incomplete 
non-spacelike geodesics, there
is a definite singular behaviour present in the \s . In such a case, a timelike
observer or a photon suddenly disappears from the \s\  after a finite amount 
of proper time or after a finite value of the affine parameter. The singularity
theorems which result from an analysis of gravitational focusing and global
properties of a \s\  prove this incompleteness property for a wide class of
\s  s under a set of rather general conditions.

The matter fields with positive energy
density affect the causality relations in a \s\  and cause focusing in the
families of timelike and null trajectories. The essential phenomena that
occurs here is that matter focuses the nonspacelike geodesics of the 
\s\  into pairs of focal points or the conjugate points. 
The rate of change of volume expansion for a given 
congruence of timelike geodesics can be written as
$${d\theta\over d\tau}=-R_{lk}V^l V^k-\textstyle{1\over3}\theta^2-2\sigma^2
+2\omega^2$$
where, for a given congruence of timelike geodesics, 
the quantities $\theta$, $\sigma$ and $\omega$ are {\it expansion},
{\it shear}, and {\it rotation} tensors are respectively.
The above equation is called the {\it Raychaudhuri equation} 
(Raychaudhuri, 1955) 
which describes the
rate of change of the volume expansion as one moves along the timelike
geodesic curves in the congruence.
We note that the second and third term on the right-hand side 
involving $\theta$ and
$\sigma$ are positive always. Consider now the term $R_{ij}V^iV^j$. By
Einstein equations this can be written as
$$R_{ij}V^i V^j= 8\pi[T_{ij}V^i V^j +\textstyle{1\over2}T]$$
The term  $T_{ij}V^i V^j$ above represents the energy density as measured by
a timelike observer with the unit tangent $V^i$, which is the four-velocity of
the observer. For all reasonable classical physical fields this energy density
is generally taken as non-negative and it is assumed that for all timelike 
vectors $V^i$ the following is satisfied
$$T_{ij}V^i V^j\ge 0$$ 
Such an assumption is called the {\it weak energy condition}.
When a suitable energy condition is satisfied, the Raychaudhuri equation
implies that the effect of matter on \s\  curvature causes a focusing effect
in the congruence of timelike geodesics due to gravitational attraction.
This, in general causes the neighbouring geodesics in the congruence to cross
each other to give rise to caustics or conjugate points. This separation 
between nearby timelike geodesics is governed by what is called 
the geodesic deviation equation,
$$ D^2Z^j= {-R^j}_{kil}V^k Z^i V^l$$
where $Z^i$ is the separation vector between  nearby geodesics of the 
congruence. Solutions of the above equation are called the {\it Jacobi fields}
along a given timelike geodesic.

There are several singularity theorems available which establish 
the non-spacelike
geodesic incompleteness for a \s\  under different sets of conditions and
applicable to different physical situations. However, the most general of 
these is the Hawking$-$Penrose theorem (Hawking and Penrose, 1970), which is 
applicable in both the collapse situation and cosmological scenario. The
main idea of the proof of such a theorem is, using the causal
structure analysis it is shown that there must be maximal length timelike
curves between certain pairs of events in the \s . Now, 
a causal geodesic which is both future and past complete must contain pairs of
conjugate points if \m\  satisfies the generic condition and an energy
condition. This is then used to 
draw the necessary contradiction to show that \m\  must be non-spacelike
geodesically
incomplete.

The inevitable existence of spacetime singularities, for 
wide classes of general models of spacetimes means that
the classical gravity necessarily gives rise to regions in the 
spacetime universe where the densities and spacetime curvatures
would really grow without any bounds, where all other physical
parameters also would diverge really.

\section{Gravitational collapse}
The existence of spacetime singularities in the Einstein 
gravity, and in all similar spacetime theories of gravitation poses 
intriguing challenges and fundamental questions in physics as well
as cosmology.

Such a phenomenon will basically arise in two physical 
scenarios in the universe, the first being the cosmology where 
such a singularity will correspond to the origin of the universe,
and secondly whenever locally a large quantity of matter and energy
collapses under the force of its own gravity. This later situation
will be effectively realized in the gravitational collapse
of massive stars in the universe, which collapse and shrink 
catastrophically under their self-gravity, when the star has
exhausted its nuclear fuel within which earlier supplied the 
internal pressure to halt the in-fall due to gravity.
We now discuss this second possibility in some detail.
 
When a massive star, more than a few solar masses, has exhausted its
internal nuclear fuel, it is believed to enter
the stage of an endless gravitational collapse without having any final
equilibrium state. According to the Einstein theory of gravitation, the
star goes on shrinking in its radius, reaching higher and higher densities.
What would be the final fate of such an object according to the general
theory of relativity? This is one of the central questions in
relativistic astrophysics and gravitation theory today. It has been
suggested that the ultra-dense object that forms as a result of collapse
could be a black hole in the space and time from which not even light rays
can escape. Alternatively, if the event horizon of gravity fails to cover the
final crunch, it could be a visible singularity which can causally interact
with the outside universe and from which emissions of light and
matter may be possible.

It is of course reasonably clear that very near such a 
spacetime singularity, the classical description that predicted
it must itself breakdown. The quantum effects associated with 
gravity are most likely to become dominant in such a regime.
These may resolve the classical singularity. However, 
we have no viable and consistent quantum theory of gravity 
available as of today despite many serious attempts, and 
therefore the issue of resolution of singularities as
produced by classical gravity remains very much open currently.

An investigation on final fate of collapse is of importance 
from both the
theoretical as well as observational point of view.
At the theoretical level, working out the collapse outcomes in
general relativity is crucial to the problem of asymptotic 
predictability, namely, whether the singularities forming
at the end point of collapse will be necessarily covered by the event
horizons of gravity. A hypothesis that remains fundamental
to the theoretical foundations of black hole physics and
its numerous astrophysical applications which have been invoked
in past decades (e.g. the area theorem for black holes, laws of black hole
thermodynamics, Hawking radiation effect, predictability; and on
observational side, accretion
of matter by black holes, massive black holes at the center of
galaxies etc),
is {\it cosmic censorship} which states the singularities of collapse
must be hidden within horizons of gravity.
On the other hand, existence of visible or naked singularities
would offer a new approach on these issues requiring modification
and reformulation of our usual theoretical conception on black holes.

To investigate this issue, dynamical
collapse scenarios have been examined in past decade or so for
many cases such as clouds composed of dust, radiation, perfect fluids,
or also of matter compositions with more general equations of
state (for references and details, 
see e.g. Joshi 2008).

\section{Black holes}
One could consider a gravitationally collapsing spherical 
massive star. We need to consider the interior solution for the 
object which will
depend on the properties of matter, equation of state, and the physical
processes taking place within the stellar interior. However, assuming
the matter to be
pressureless dust allows to solve the problem analytically, providing
many important insights. Here the energy-momentum tensor
is given by $T^{ij}= \rho u^iu^j$, and one needs to solve the Einstein
equations for the spherically symmetric metric.
This determines the metric potentials,
and the interior geometry of the collapsing dust ball is given by,
$$ds^2= -dt^2+ R^2(t)\left[ {dr^2\over 1-r^2}+ r^2 d\Omega^2 \right]$$
where $d\Omega^2=d\theta^2+ sin^2\theta d\phi^2$ is the metric on two-sphere.
The geometry outside is vacuum Schwarzschild space-time. The interior
geometry of the dust cloud matches at the
boundary $r=r_b$ with the exterior Schwarzschild space-time.

\begin{center}
\leavevmode\epsfysize=3.5 in\epsfbox{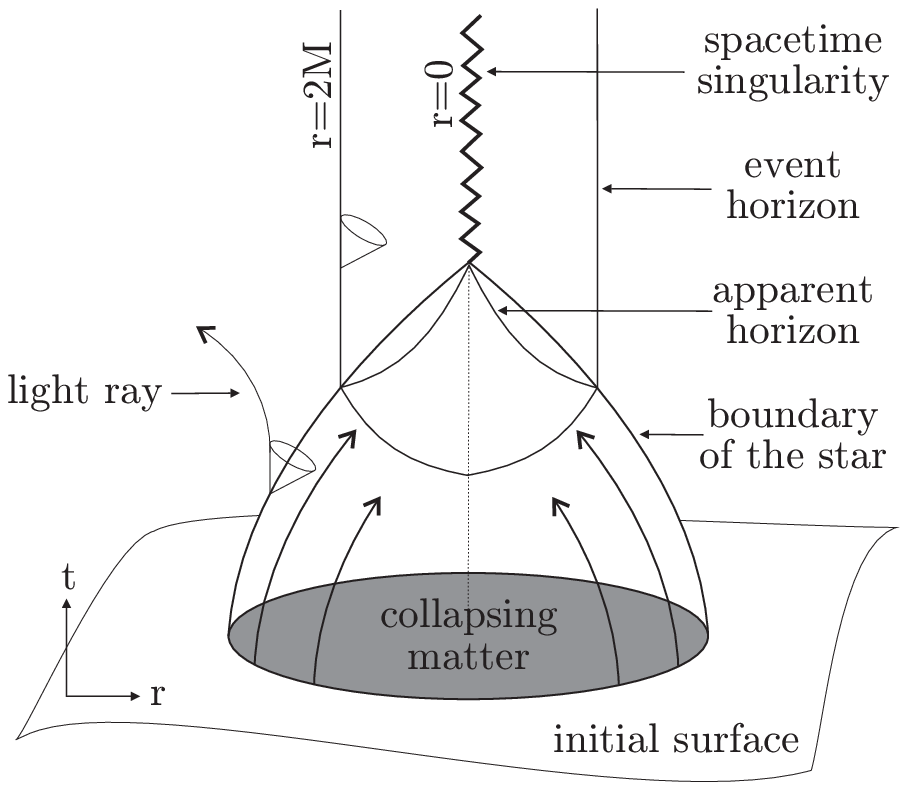}
\end{center}
\noindent {\small Fig 1. Spherically
symmetric homogeneous dust collapse where the final outcome is 
black hole formation in the spacetime.}

The basic features of such a collapsing, spherical, homogeneous dust
cloud configuration are given in Fig. 1.  The collapse is
initiated when the star surface is outside its Schwarzschild radius $r = 2m$,
and a light ray emitted from the surface of the star can escape to infinity.
However, once the star has collapsed below $r = 2m$, a black hole region
of no escape develops in the space-time, bounded by the event horizon
at $r =2m$.  Any point in this empty region
represents a trapped surface (which is a two-sphere in space-time)
in that both the outgoing and ingoing families of null geodesics emitted
from this point converge and hence no light comes out of this region.
Then, the collapse to an infinite density and
curvature singularity at $r =0$ becomes inevitable
in a finite proper time as measured
by an observer on the surface of the star. The black hole region in the
resulting vacuum Schwarzschild geometry is given by $0 < r < 2m$,
the event horizon being the outer boundary.
On the event horizon, the radial outwards
photons stay where they are, but all the rest are dragged inwards. No
information from this black hole can propagate outside $r=2m$ to
observers far away.
We thus see that the collapse gives rise to a
black hole in the space-time which covers the
resulting space-time singularity. The ultimate fate of the star
undergoing such a collapse is then an infinite curvature singularity
at $r = 0$, which is completely hidden within the black hole. No emissions or
light rays  from the singularity could go out to observer at infinity and the
singularity is causally disconnected from the outside space-time.

\section{Cosmic censorship}
The question now is whether one could generalize these conclusions
on the occurrence of a spacetime singularity in collapse
and black hole formation for more general matter fields, and
for non-spherical situations. 

While we know that the occurrence of the singularity itself 
is stable to small perturbations in the initial data, there is no 
proof available that such a singularity will continue 
to be hidden within a black hole and remain causally disconnected
from outside observers, even when the collapse is not spherical
or when the matter does not have the form of exact homogeneous dust.

Thus, in order to generalize the notion of black holes to gravitational
collapse situations other than exact spherically symmetric homogeneous
dust case, it becomes necessary to rule out such naked or visible
singularities by means of an explicit assumption.  This is stated as the
{\it cosmic censorship hypothesis}, which essentially states that if
$S$ is a partial Cauchy surface from which collapse commences, then there
are no naked singularities to the future of $S$, that is, which could
be seen from the future null infinity. This is true for the spherical
homogeneous dust collapse, where the resulting spacetime is future
asymptotically predictable and the censorship holds. 
Thus, the breakdown
of physical theory at the spacetime singularity does not disturb prediction
in future for the outside asymptotically flat region.
What will be the corresponding scenario for other collapse situations,
when inhomogeneities, non-sphericity etc are allowed for? 
It is clear that the assumption of censorship in a suitable form is crucial
to basic results in black hole physics.  In fact, when one considers the
gravitational collapse in a generic situation, the very existence of
black holes requires this hypothesis.

If one is to establish the censorship by means of a rigorous proof,
that of course requires a much more precise formulation of the hypothesis.
The statement that result of a complete gravitational collapse must
always be a black hole and not a naked singularity, or all singularities of
collapse must be hidden in black holes, causally disconnected from observers
at infinity, is not rigorous enough. This is because, under completely general
circumstances, the censorship or asymptotic predictability is false as one
could always choose a space-time manifold with a naked singularity which would
be a solution to Einstein's equations if we define
$T_{ij} \equiv (1/8\pi)G_{ij}$.
In fact, as far as the cosmic censorship
hypothesis is concerned, it is a major problem in itself to find a
satisfactory and mathematically rigorous  formulation of what is physically
desired to be achieved.  Developing a suitable formulation would probably be
a major advance towards the solution of the main problem. It should be
noted that presently no general proof
is available for any suitably formulated version of the weak
censorship.

\section{Naked singularities}
This situation leads us to conclude that the first
and foremost task is to carry out a detailed and careful
examination of various gravitational collapse scenarios
to examine them for their end states.
It is clear from these considerations that
we still do not have sufficient data and information
available on the various possibilities for gravitationally
collapsing configurations so as to decide one way or other on the
issue of censorship hypothesis. What appears really necessary is a detailed
investigation of different collapse scenarios, and to examine the
possibilities arising, in order to have
insights into the issue of the final fate of gravitational collapse.
With such a purpose, several gravitational collapse scenarios
involving different forms of matter have been investigated
to understand better the final fate of collapse.

Since we are interested in collapse,
we require that the space-time contains a regular initial spacelike
hypersurface on which the matter fields, as represented by the stress-energy
tensor $T_{ij}$, have a compact support and all physical quantities are
well-behaved on this surface. Also, the matter should
satisfy a suitable energy condition and the Einstein equations are
satisfied. We say that the space-time contains a naked singularity
if there is a future directed non-spacelike curve which reaches a far
away observer or infinity in future, and in the past it terminates
at the singularity.

As an immediate generalization of the Oppenheimer-Snyder-Datt
homogeneous dust collapse, one could consider the collapse of 
inhomogeneous dust and examine the nature and structure of resulting 
singularity with special reference to censorship, and the occurrence 
of black holes and naked singularities. The main motivation to 
discuss this situation is this provides a clear picture in 
an explicit manner of what is possible in
gravitational collapse. One could ask how are the conclusions 
given above for homogeneous collapse are modified when the 
inhomogeneities of matter distribution are taken into account.
Clearly, it is important to include effects of inhomogeneities because
typically a realistic collapse would start from a very inhomogeneous
initial data with a centrally peaked density profile.

This problem was investigated in detail using the 
Tolman-Bondi-Lemaitre models, which describe gravitational collapse 
of an inhomogeneous spherically symmetric dust cloud
(Joshi and Dwivedi 1993).  
This is an infinite dimensional family of asymptotically flat 
solutions of Einstein equations, which is matched to the Schwarzschild 
spacetime outside the boundary of the collapsing star. The 
Oppenheimer-Snyder-Datt model is a special case of this class
of solutions.

It is seen that the introduction of inhomogeneities
leads to a rather different picture of gravitational collapse.  
The metric for spherically
symmetric collapse of inhomogeneous dust, in comoving
coordinates $(t, r, \theta,\phi)$, is given by,
$$ ds^2= -dt^2+{R'^2\over1+f}dr^2+R^2(d\theta^2+sin^2\theta\, d\phi^2)$$

$$T^{ij}=\epsilon \delta^i_t \delta^j_t,\quad \epsilon=\epsilon(t,r)={F'
\over R^2R'}$$

where $T^{ij}$ is the stress-energy tensor, $\epsilon$ is
the energy density, and $R$ is a function of both $t$ and $r$ given by
$$ \dot R^2={F\over R}+f $$
Here the dot and prime denote partial derivatives with respect
to the parameters $t$ and $r$ respectively. As we are considering collapse,
we require $\dot R(t,r)<0.$ The quantities $F$ and $f$ are arbitrary
functions of $r$ and $4\pi R^2(t,r)$ is the proper area of the  mass shells.
The area of such a shell at $r=\hbox{const.}$ goes to zero when $R(t,r)=0$.
For gravitational collapse situation,
we take $\epsilon$ to have compact support
on an initial spacelike hypersurface and the space-time can be matched
at some $r=\hbox{const.}=r_c$ to the exterior Schwarzschild field
with total Schwarzschild mass $m(r_c)=M$ enclosed within the dust ball
of coordinate radius of $r=r_c$. The apparent horizon in the interior
dust ball lies at $R=F(r)$.

Using this framework, the nature of the singularity $R=0$ can
be examined. In particular, the problem of nakedness or otherwise of
the singularity can be reduced to the existence of real, positive roots
of an algebraic equation $V(X)=0$, constructed out of 
the free functions $F$ and
$f$ and their derivatives [11], which constitute the initial data of
this problem. If the equation $V(X)=0$ has a real 
positive root, the singularity
could be naked. In order to be the end point of null geodesics at least
one real positive value of $X_0$ should satisfy  the above.
Clearly, if no real positive root of the above is found, the singularity
is not naked. It should be noted that
many real positive roots of the above equation may exist which give the
possible values of tangents to the singular null geodesics
terminating at the singularity. Suppose now $X=X_0$ is 
a simple root to $V(X)=0$.
To determine whether $X_0$ is realized as a tangent along any outgoing
singular geodesics to give a naked singularity, one can integrate the
equation of the radial null geodesics in the form $r=r(X)$
and it is seen that there is always at least one null geodesic terminating
at the singularity $t=0,r=0$, with $X=X_0$. In addition there would be
infinitely many integral curves as well, depending on the values of the
parameters involved, that terminate at the singularity.
It is thus seen that the existence of a positive real root of
the equation $V(X)=0$ is a
necessary and sufficient condition for the singularity to be
naked. Finally, to determine the curvature strength of the naked singularity
at $t=0$, $r=0$, one may analyze the quantity
$ k^2 R_{ab}K^aK^b$ near the singularity. Standard analysis shows that
the strong curvature condition is satisfied, in that the above
quantity remains finite in the limit of approach to the singularity.

\section{General collapse scenarios}
The assumption of vanishing pressures,
which could be important in the final stages of the collapse, may be
considered as the limitation of dust models.
On the other hand, it is also argued sometimes that in the final stages
of collapse, the dust equation of state could be relevant
and at higher and higher densities the matter may behave
more like dust. Further, if there are no large negative pressures
(as implied by the validity of the energy conditions), then the pressure
also might contribute gravitationally in a positive manner
to the effect of dust and may not alter the conclusions.

In any case, it is important to consider collapse situations 
with non-zero pressures and with reasonable equations of state.
Pressures may play an important role for the later
stages of collapse and one must investigate the possibility if pressure
gradients could prevent the occurrence of naked singularity.
These issues have been examined namely, the existence,
the termination of non-spacelike geodesic families, and the
strength of such a singularity for collapse with non-zero 
pressure. The results could be summarized as follows. If in a self-similar
collapse with pressure, a single null radial geodesic escapes the 
singularity, then an entire family of non-spacelike geodesics 
would also escape provided the positivity of energy density
is satisfied as above.

Actually, gravitational collapse models with a general form
of matter, together with those such as directed radiation, dust,
perfect fluids etc imply some general pattern emerging about the final
outcome of gravitational collapse. Basically it follows that 
the occurrence of naked singularity is basically
related to the choice of initial data to the Einstein field equations,
and would therefore occur from  regular initial data within
the general context considered, subject to the matter satisfying
weak energy condition. It appears that the occurrence of naked 
singularity or a black hole is more a problem of choice of the 
initial data for field equations rather than that of the form 
of matter or the equation of state.
This has important implication for cosmic censorship in that in order to
preserve the same one has to avoid all such regular initial data causing
naked singularity, and hence a deeper understanding of the initial data
space is required in order to determine such initial data and the kind
of physical parameters they would specify. This would, in other words,
classify the range of physical parameters to be avoided for a particular
form of matter. More importantly, it would also pave the way for the
black hole physics to use only those ranges of allowed parameter values which
would produce black holes, thus putting black hole physics on a more
firm footing.

What will be the final fate of gravitational collapse which is not
spherically symmetric? The main phases of spherical collapse of a
massive star would be typically instability,
implosion of matter, and subsequent formation of an event horizon and
a space-time singularity of infinite density and curvature with infinite
gravitational tidal forces. This singularity may or may not be fully
covered by the horizon as we have discussed above.
Again, small perturbations over the spherically symmetric situation
would leave the situation unchanged in the sense that an event horizon
will continue to form in the advanced stages of the collapse.

The question then is, do horizons still form when the fluctuations 
from the spherical symmetry are high and the collapse is highly non-spherical?
It was shown by 
Thorne (1972), 
that when there is no spherical
symmetry, the collapse of infinite cylinders do give rise to naked
singularities in general relativity, which are not covered by horizons.
This situation motivated Thorne to propose the following {\it hoop conjecture}
for finite systems in an asymptotically flat space-time, which
characterizes the final fate of non-spherical collapse: The
horizons of gravity form when and only when a mass $M$ gets compacted in a
region whose circumference in {\it every} direction obeys
$ {\cal C}\le 2\pi(2GM/c^2)$. Thus, unlike the cosmic censorship conjecture,
the hoop conjecture does not rule out {\it all} the naked singularities
but only makes a definite assertion on the occurrence of the event
horizons in gravitational collapse. We also note that the hoop conjecture
is concerned with the formation of event horizons, and not with naked
singularities. Thus, even when event horizons form, say for example in
the spherically symmetric case, it does not rule out the existence of
naked singularities, i.e. it does not imply that such horizons must
always cover the singularities.

Apart from such numerical simulations, some analytic treatments of
aspherical collapse are also available. For example, the aspherical
Szekeres models for irrotational dust without any Killing vectors,
generalizing the spherical Tolman-Bondi-Lemaitre collapse, were 
studied, to deduce the existence of strong curvature central naked
singularities.
While this indicates that naked singularities are not necessarily confined
to spherical symmetry, it must be noted that dynamical evolution
of a non-spherical collapse still remains a largely uncharted territory.

We note here that the {\it genericity} and {\it stability} 
of the collapse outcomes, in terms of
black holes and naked singularities need to be understood 
carefully and in further detail. It is by and large well-accepted
now, that the general theory of relativity does allow and
gives rise to both black holes and naked singularities as final
fate of a continual gravitational collapse, evolving from a
regular initial data, and under reasonable physical conditions.
What is not fully clear as yet is the distribution
of these outcomes in the space of all allowed outcomes
of collapse. The collapse models discussed above and considerations
we gave here would be of some help in this direction, and may
throw some light on the distribution of black holes and
naked singularity solutions in the initial data space.

The important question then is the genericity and stability of such
naked singularities arising from regular initial data. Will the initial
data subspace, which gives rise to naked singularity as end state
of collapse, have zero measure in a suitable sense? In that case, one
would be able to reformulate more suitably the censorship hypothesis,
based on a criterion that naked singularities
could form in collapse but may not be generic.

It is natural to ask here, what is really the physics that          
causes a naked singularity to develop
in collapse, rather than a black hole? We need to know how at all
particles and energy are allowed to escape from extremely strong
gravity fields. We have examined this issue in some detail to bring
out the role of inhomogeneities and space-time shear to achieve this
towards distorting the geometry of horizons forming in collapse.
In Newtonian gravity, it is only the matter density that determines the
gravitational field. In Einstein theory, however, density is just one
attribute of the overall gravitational field, and the various
curvature components and scalar quantities play an equally important
role to dictate what the overall nature of the field is.
What our results show is, once the density is inhomogeneous          
or higher at the center of
collapsing star, this rather naturally delays the trapping of light
and matter during collapse, which can in principle escape away. This
is a general relativistic effect wherein even if the densities are
very high, paths are created for light or matter to escape due to
inhomogeneously collapsing matter fields, and  these physical 
features naturally lead to a naked singularity
formation rather than a black hole end state. It is the amount of
inhomogeneity that counts to distort the horizons. If it is very
small, below a critical limit, a black hole will form, but with sufficient
inhomogeneity trapping is delayed to cause
a naked singularity. This criticality again comes out in the Vaidya
class of radiation collapse models, where it is the rate of collapse,
that is how fast or slow the cloud is collapsing, that determines the
black hole or naked singularity formation.

\section{Distinguishing black holes and naked singularities observationally}
It is clear that the black hole and naked singularity
outcomes of a complete gravitational collapse for a massive
star are very different from each other physically, and
would have quite different observational signatures.
In the naked singularity case, if it occurs in nature,
we have the possibility to observe the physical effects
happening in the vicinity of the ultra dense regions that form
in the very final stages of collapse. However, in a black
hole scenario, such regions are necessarily hidden
within the event horizon of gravity.

There have been attempts where researchers explored 
physical applications and implications of the naked singularities 
(see e.g. 
Joshi and Malafarina 2011
and references in there).
If we could find astrophysical applications of the models
that predict naked singularities as collapse final fate, and possibly 
try to test the same through observational methods and the signatures 
predicted, that could offer a very interesting avenue to get 
further insight into the problem as a whole. An attractive recent 
possibility in that connection is to explore the naked singularities 
as possible particle accelerators 
(Patil and Joshi 2011).

Also, the accretion discs around a naked singularity,
wherein the matter particles are attracted towards or repulsed
away from the singularities with great velocities could provide
an excellent venue to test such effects and may lead to
predictions of important observational signatures to
distinguish the black holes and naked singularities in
astrophysical phenomena. The question of what observational signatures 
would then emerge and distinguish the black holes from naked singularities
is then necessary to be investigated, and we must explore
what special astrophysical consequences the latter may have.

Where could the observational signatures of naked singularities
lie? If we look for the sign of singularities such as the ones that
appear at the end of collapse, we have to consider explosive and high
energy events. In fact such models expose the ultra-high density
region at the time of formation of the singularity while the outer
shells are still falling towards the center. In such a case,
shockwaves emanating from the superdense region at scales smaller than
the Schwarzschild radius (that could be due to quantum effects or
repulsive classical effects) and collisions of particles near the
Cauchy horizon could have effects on the outer layers. These would be
considerably different from those appearing during the formation of 
a black hole. If, on the other hand, we consider singularities such as the
superspinning Kerr solution we can look for different kinds of
observational signatures. Among these the most prominent features deal
with the way the singularity could affect incoming particles, either
in the form of light bending, such as in gravitational lensing,
particle collisions close to the singularity, or properties of
accretion disks.

Essentially we ask whether we could test censorship using 
astronomical observations.
With so many high technology power missions to observe the
cosmos, can we not just observe the skies carefully to determine
the validity or otherwise of the cosmic censorship?
In this connection, several proposals to measure the mass and 
spin ratio for compact objects and for the galactic center 
have been made by different researchers. 
In particular, using pulsar observations it is suggested
that gravitational waves and the spectra of X-rays binaries 
could test the rotation parameter for the center of
our galaxy. Also, the shadow cast by the compact object 
can be used to test the same in stellar mass objects, 
or X-ray energy spectrum emitted by the
accretion disk can be used. Using certain observable properties 
of gravitational lensing that depend upon rotation
is also suggested (for references, see
Joshi and Malafarina, 2011).

The basic issue here is that of sensitivity, namely how
accurately and precisely can we measure and determine these
parameters. A number of present and future astronomical
missions could be of help. One of these is the Square-Kilometer
Array (SKA) radio telescope, which will offer a possibility
here, with a collecting area exceeding a factor of hundred
compared to existing ones. The SKA astronomers point out they
will have the sensitivity desired to measure the required quantities
very precisely to determine the vital fundamental issues in
gravitation physics such as the cosmic censorship, and
to decide on its validity or otherwise.
Other missions that could in principle provide a huge
amount of observational data are those that are currently hunting
for the gravitational waves. Gravitational wave astronomy
has yet to claim its first detection of waves, nevertheless
in the coming years it is very likely that the first
observations will be made by the experiments
such as LIGO and VIRGO that are currently still below
the threshold for observation. Then gravitational wave
astronomy will become an active field with possibly large
amounts of data to be checked against theoretical
predictions and it appears almost certain that this
will have a strong impact on open theoretical issues
such as the Cosmic Censorship problem.

There are three different kinds of observations that one could
devise in order to distinguish a naked singularity from a black hole.
The first one relies on the study of accretion disks. The accretion
properties of particles falling onto a naked singularity would be very
different from those of black hole of the same mass (see for example 
(Pugliese et al, 2010; 
Joshi, Malafarina and Ramesh Narayan, 2011),
and the resulting accretion disks would also be observationally different.
The properties of accretion disks have been studied
in terms of the radiant energy, flux and luminosity, in a Kerr-like
geometry with a naked singularity, and the differences from a black 
hole accretion disk have been investigated.
Also, the presence of a naked singularity gives rise to powerful
repulsive forces that create an outflow of particles from the
accretion disk on the equatorial plane.
This outflow that is otherwise not present in the black hole case,
could be in principle distinguished from the jets of particles
that are thought to be ejected from black hole's polar region and
which are due to strong electromagnetic
fields. Also, when charged test particles are considered
the accretion disk's properties
for the naked singularity present in the Reissner-Nordstrom spacetime
are seen to be observationally different from those
of black holes.

The second way of distinguishing black holes from naked singularities relies
on gravitational lensing. It is argued that when the spacetime does not
possess a photon sphere, then the lensing features of light passing close to
the singularity will be observationally different from those of a black hole.
This method, however, does not appear to be very effective when a
photon sphere is present in the spacetime.
Assuming that a Kerr-like solution of Einstein equations with
massless scalar field exists at the center of galaxies,
its lensing properties are studied
and it was found that there are effects due to the presence of
both the rotation and scalar field that would affect the behavior of
the bending angle of the light ray, thus making those objects
observationally different from black holes.

Finally, a third way of distinguishing black holes from naked singularities
comes from particle collisions and particle acceleration in the vicinity of
the singularity. In fact, it is possible that the repulsive effects due to
the singularity can deviate a class of infalling particles, making
these outgoing eventually. These could then collide with some
ingoing particle, and the energy of collision
could be arbitrarily high, depending on the impact parameter of the
outgoing particle with respect to the singularity. The net effect is
thus creating a very high energy collision that resembles that of
an immense particle accelerator and that would be impossible in
the vicinity of a Kerr black hole.

\section{Predictability and other cosmic puzzles}
What then is the status of naked singularities versus censorship 
today? Can cosmic censorship survive in some limited and specialized
form, and firstly, can we properly formulate it after all these 
studies in recent years on gravitational collapse?
While this continues to be a major cosmic puzzle, recent 
studies on formation of naked singularities as collapse end states 
for many realistic models have 
brought to forefront some of the most intriguing basic questions, both
at classical and quantum level, which may have significant physical
relevance. Some of these are: Can the super ultra-dense regions
forming in a physically realistic collapse of a massive star be
visible to far away observers in space-time? Are there any observable
astrophysical consequences? What is the causal structure of space-time
in the vicinity of singularity as decided by the internal dynamics of
collapse which evolves from a regular initial data at an initial
time?  How early or late the horizons will actually develop in a
physically realistic gravitational collapse, as determined by the 
astrophysical conditions within the star? When a naked 
singularity forms, is it possible to observe the quantum gravity 
effects taking place in the ultra-strong gravity regions? Can one 
possibly envisage a connection to observed ultra-high energy 
phenomena such as cosmic gamma ray bursts?

A continuing study of collapse phenomena 
within a general and physically realistic framework may be the only
way to answers on some of these issues. This could lead us to 
novel physical insights and possibilities emerging out of the 
intricacies of gravitational force and nature of gravity, as emerging 
from examining the dynamical evolutions as allowed by 
Einstein equations.

Apart from its physical relevance, the collpase phenomena 
also have profound philosophical implications such as on the issue 
of predictability in the universe. We summarize below a few arguments, 
for and against it in the classical general relativity.

It is some times argued that breakdown of censorship means violation
of predictability in spacetime,because we have no direct handle 
to know what a naked singularity may radiate and emit unless we
study the physics in such ultra-dense regions. One would not be able
then to predict the universe in the future of a given epoch of time as
would be the case, for example, in the case of the Schwarzschild black
hole that develops in Oppenheimer-Snyder collapse.

A concern usually expressed is that if naked
singularities occurred as the final fate of gravitational
collapse, that would break the predictability in the
spacetime, because the naked singularity is characterized by the
existence of light rays and particles that emerge from the
same. Typically, in all the collapse models discussed
above, there is a family of future directed non-spacelike
curves that reach external observers, and when extended
in the past these meet the singularity.
The first light ray that comes out from the singularity
marks the boundary of the region that can be predicted
from a regular initial Cauchy surface in the spacetime,
and that is called the Cauchy horizon for the spacetime.
The causal structure of the spacetime would differ
significantly in the two cases, when there is a Cauchy
horizon and when there is none.

In general relativity, a given `epoch' of time is sometimes          
represented by a spacelike surface, which is three-dimensional space. 
For example, in the standard Friedmann models of cosmology, there
is such an epoch of simultaneity, from which the universe evolves in
future, given the physical variables and initial data on this
surface. The Einstein equations govern this evolution of universe, and
there is thus a predictability which one would expect to hold in a
classical theory. The concern that is expressed at times is one would
not be able to predict in the future of naked singularity, and that
unpredictable inputs may emerge from the same.

The point here is, given a regular initial data on a
spacelike hypersurface, one would like to predict the future
and past evolutions in the spacetime for all times (see for example, 
Hawking and Ellis 1973).
Such a requirement is termed as the {\it global hyperbolicity}
of the spacetime. A globally hyperbolic spacetime is a fully
predictable universe, it admits a {\it Cauchy surface}, 
which is a three dimensional spacelike surface the data
on which can be evolved for all times in the past as well
as in future. Simple enough spacetimes such as the Minkowski
or Schwarzschild are globally hyperbolic, but the Reissner-Nordstrom
or Kerr geometries are not globally hyperbolic. For further
details on these issues, we refer to 
(Joshi, 2008).

The key role that the event horizon of a black hole plays
is that it hides the super-ultra-dense region formed in collapse
from us. So the fact that we do not understand such regions
has no effect on our ability to predict what happens
in the universe at large. But if no such horizon exists, then
the ultra-dense region might, in fact, play an important and
even decisive role in the rest of the universe, and our ignorance
of such regions would become of more than merely academic
interest.

Yet such an unpredictability is common in general relativity,
and not always directly related to censorship violation. Even black
holes themselves need not fully respect predictability when
they rotate or have some charge. For example, if we drop an
electric charge into an uncharged black hole, the spacetime geometry
radically changes and is no longer predictable from a regular
initial epoch of time. A charged black hole admits a naked
singularity which is visible to an observer within the horizon,
and similar situation holds when the black hole is rotating.
There is an important debate in recent years, if one could 
over-charge or over-rotate a black hole so that the singularity
visible to observers within the horizon becomes visible to external
far away observers too. 

Another point is, if such a black hole was big enough on a
cosmological scale, the observer within the horizon could
survive in principle for millions of years happily without
actually falling into the singularity, and would thus be able
to observe the naked singularity for a long time. Thus,
only purest of pure black holes with no charge or rotation at
all respect the full predictability, and all other physically
realistic ones with charge or rotation actually do not.
As such, there are many models of the universe in cosmology
and relativity that are not totally predictable from a given
spacelike hypersurface in the past. In these universes, the
spacetime cannot be neatly separated into space and time foliation
so as to allow initial data at a given moment of time to
fully determine the future.

In our view, the real breakdown of predictability is the
occurrence of spacetime singularity itself, which indicates
the true limitation of the classical gravity theory. It does
not matter really whether it is hidden within an event horizon
or not. The real solution of the problem would then be the
resolution of singularity itself, through either a quantum
theory of gravity or in some way at the classical level
itself.

Actually, the cosmic censorship way to predictability,
that of `hiding the singularity within a black hole', and
then thinking that we restored the spacetime predictability
may not be the real solution, or at best it may be only a partial
solution to the key issue of predictability in spacetime universes.
In fact, it may be just shifting the problem elsewhere, and some
of the current major paradoxes faced by the black hole physics
such as the information paradox, the various puzzles regarding
the nature of the Hawking radiation, and other issues could
as well be a manifestation of the same.

No doubt, the biggest argument in support of censorship would 
be that it would justify and validate the extensive formalism and 
laws of black hole physics and its astrophysical applications 
made so far.  Censorship has been the foundation
for the laws of black holes such as the area theorem and others,
and their astrophysical applications. But these are not free
of major paradoxes.
Even if we accept that all massive stars 
would necessarily turn into black holes, this still creates 
some major physical paradoxes. Firstly, all the matter entering 
a black hole must of necessity collapse into a space-time singularity of
infinite density and curvatures, where all known laws of physics 
break down, which is some kind of instability at the classical 
level itself. This was a reason why many gravitation theorists 
of 1940s and 1950s objected to black hole formation, and Einstein 
also repeatedly argued against such a final fate of a collapsing star, 
writing a paper in 1939 to this effect.
Also, as is well-known and has been widely discussed in the 
past few years, a black hole, by potentially destroying information, 
appears to contradict the basic principles of quantum theory. In that
sense, the very formation of a black hole itself with a
singularity within it appears to come laden with inherent
problems. It is far from clear how one would resolve these
basic troubles even if censorship were correct.

In view of such problems with the black hole paradigm,
a possibility worth considering is the delay or avoidance of
horizon formation as the star collapses under gravity. This
happens when collapse to a naked singularity takes place, namely,
where the horizon does not form early enough or is avoided.
In such a case, if the star could radiate away most of its mass
in the late stages of collapse, this may offer a way out of
the black hole conundrum, while also resolving the singularity
issue, because now there is no mass left to form the singularity.

What this means is, such an `unpredictability' is somewhat
common in general relativity. For example, if we drop a slight charge 
in a Schwarzschild black hole, the spacetime geometry
completely changes into that of a charged black hole that is no longer
predictable in the above sense. Similar situation holds when the 
black hole is rotating. In fact, there are very many models of universe
in use in relativity which are not `globally hyperbolic', that is, not
totally predictable in the above sense where space and time are 
neatly separated so as to allow initial data to fully determine future
for all times.

In any case, a positive and useful feature that has emerged
from work on collapse models 
so far is, we already have now several important constraints 
for any possible formulation of censorship. It is seen that several 
versions of censorship proposed earlier would not hold,
because explicit counter-examples are available now.
Clearly, analyzing gravitational collapse
plays a crucial role here. Only if we understand clearly why naked
singularities do develop as collapse end states in many realistic
models, there could emerge any pointer or lead to any practical 
and provable version of censorship.

Finally, it may be worth noting that even if the problem of singularity 
was resolved somehow,
possibly by invoking quantum gravity which may smear the singularity, 
we still have to mathematically formulate and prove the black 
hole formation assuming an appropriate censorship principle, which 
is turning out to be most difficult task with no sign of resolve. 
As discussed, the detailed collapse calculations of recent years 
show that the final fate of a collapsing star could be a naked
singularity in violation to censorship.  Finally, as is well-known 
and widely discussed by now, a black hole creates the information 
loss paradox, violating unitarity and making contradiction with 
basic principles of quantum theory. It is far from clear how one 
would resolve these basic troubles even if censorship were correct.

\section{A Lab for quantum gravity--Quantum Stars?}
It is believed that when we have a reasonable and complete 
quantum theory of gravity available, all spacetime singularities, 
whether naked or those hidden inside black holes, will be 
resolved away. As of now, it remains an open question if the quantum
gravity will remove naked singularities.
After all, the occurrence of spacetime singularities
could be a purely classical phenomenon, and
whether they are naked or covered should not be relevant,
because quantum gravity will possibly remove them
all any way. 
It is possible that in a suitable quantum gravity theory
the singularities will be smeared out, though this has been
not realized so far.

In any case, the important and real issue is,
whether the extreme strong gravity regions formed due
to gravitational collapse are visible to faraway observers
or not. It is quite clear that the gravitational collapse
would certainly proceed classically, at least till the
quantum gravity starts governing and dominating the
dynamical evolution at the scales of the order
of the Planck length, {\it i.e.} till the extreme gravity
configurations have been already developed due to
collapse. The point is, it is the visibility or
otherwise of such ultra-dense regions that is under
discussion, whether they are classical or quantum
(see Fig.2).


\begin{center}
\leavevmode\epsfysize=3.5 in\epsfbox{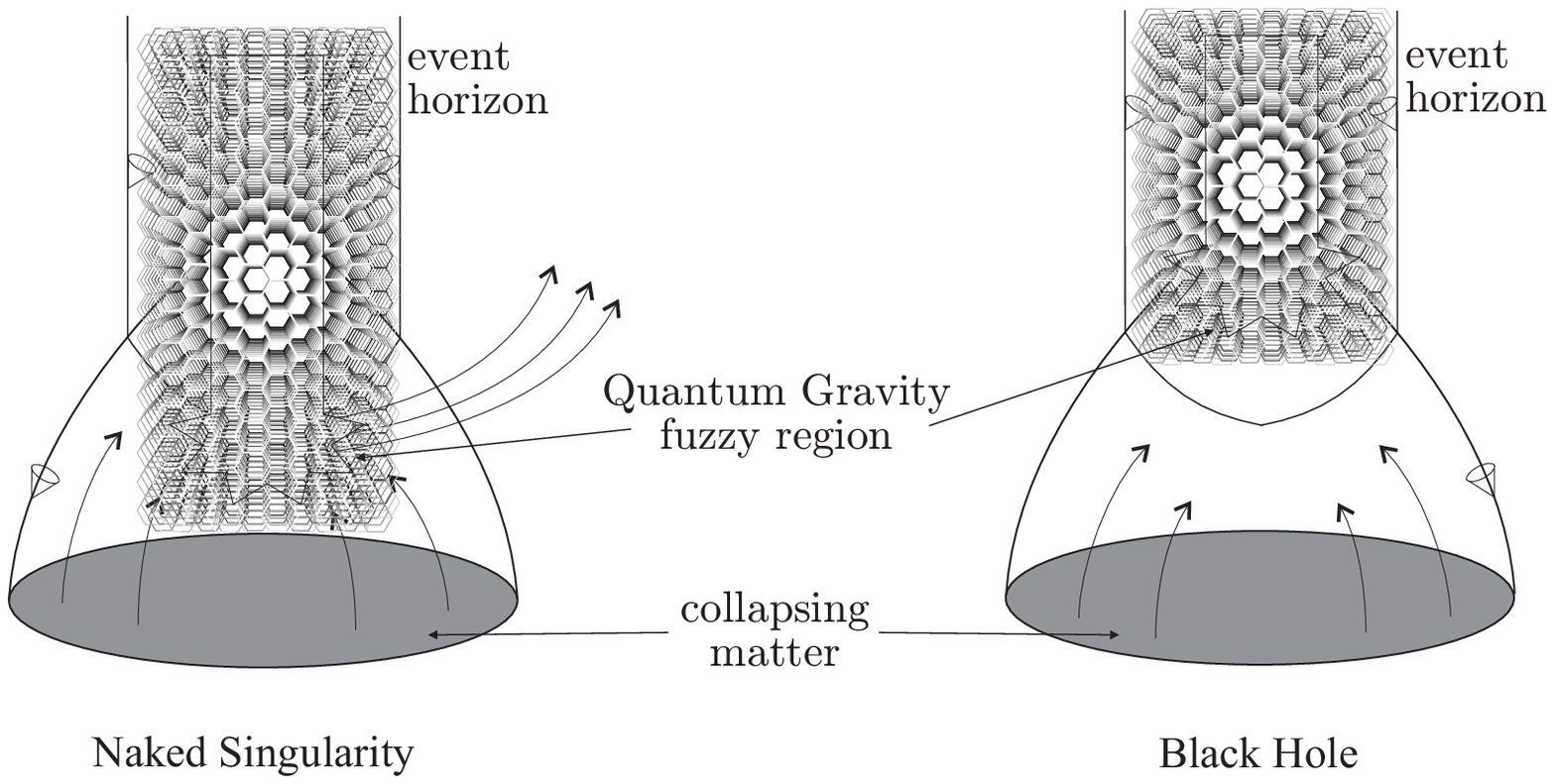}
\end{center}
\noindent {\small Fig 2: The naked singularity may be resolved by
the quantum gravity effects but the ultra-strong
gravity region that developed in gravitational collapse
will still be visible to external observers
in the universe.}

What is important is, classical gravity implies
necessarily the existence of ultra-strong gravity
regions, where both classical and quantum gravity come into
their own. In fact, if naked singularities do develop in
gravitational collapse, then in a literal sense we come
face-to-face with the laws of quantum gravity, whenever
such an event occurs in the universe.

In this way, the gravitational collapse phenomenon
has the potential to provide us with a possibility of
actually testing the laws of quantum gravity.
In the case of a black hole developing in the
collapse of a finite sized object such as a massive star,
such strong gravity regions are necessarily hidden
behind an event horizon of gravity, and this would be
well before the physical conditions became extreme
near the spacetime singularity.
In that case, the quantum effects, even if they caused
qualitative changes closer to singularity, will be
of no physical consequences as no causal
communications are then allowed from such regions. On
the other hand, if the causal structure were that
of a naked singularity, then the communications from
such a quantum gravity dominated extreme curvature
ball would be visible in principle. This will be so
either through direct physical processes near a
strong curvature naked singularity, or via the
secondary effects, such as the shocks produced in
the surrounding medium.
It is possible that a spacetime singularity basically
represents the incompleteness of the classical theory and
when quantum effects are combined with the gravitational
force, the classical singularity may be resolved.

Therefore, more than the existence of a naked singularity,
the important physical issue is whether the extreme
gravity regions formed in the gravitational collapse of a
massive star are visible to external observers in the universe.
An affirmative answer here would mean that such a collapse
provides a good laboratory to study quantum gravity effects in
the cosmos, which may possibly generate clues for an as yet
unknown theory of quantum gravity. Quantum gravity theories
in the making, such as the string theory or loop quantum
gravity in fact are badly in need of some kind of an observational
input, without which it is nearly impossible to constrain
the plethora of possibilities.

We could say quite realistically that a laboratory similar
to that provided by the early universe is created in the collapse
of a massive star. However, the big bang, which is also
a naked singularity in that it is in principle visible to all
observers, happened only once in the life of the universe
and is therefore a unique event. But a naked singularity of
gravitational collapse could offer an opportunity to explore
and observe the quantum gravity effects every time a massive
star in the universe ends its life.

The important questions one could ask are: If in realistic
astrophysical situations the star terminates as a naked singularity,
would there be any observable consequences which reflect the
quantum gravity signatures in the ultra-strong gravity region?
Do naked singularities have physical properties different
from those of a black hole? Such questions underlie our
study of gravitational collapse.

In view of recent results on gravitational collapse, and 
various problems with the black hole paradigm, a
possibility worth considering is the delay or avoidance of
horizon formation as the star evolves collapsing under
gravity. This happens when collapse to a naked singularity
takes place, where the horizon does not form early enough or
is avoided.  In such a case, in the late stages of collapse
if the star could radiate away most of its mass,
then this may offer a way out of the black hole conundrum, while also
resolving the singularity issue, because now there is no mass left to
form the curvature singularity. The purpose is to resolve the black
hole paradoxes and avoid the singularity, either visible or within a
black hole, which actually indicates the breakdown of physical theory.
The current work on gravitational collapse suggests possibilities in
this direction.

In this context, we
considered a cloud that collapsed to a naked singularity
final state, and introduced loop quantum gravity effects
(Goswami, Joshi and Singh, 2006).
It turned out that the quantum effects generated an extremely
powerful repulsive force within the cloud. Classically the cloud would
have terminated into a naked singularity, but quantum effects
caused a burstlike emission of matter in the very last phases of
collapse, thus dispersing the star and dissolving the naked
singularity. The density remained finite and the spacetime
singularity was eventually avoided.  One could expect this to
be a fundamental feature of other quantum gravity theories
as well, but more work would be required to confirm such a
conjecture.

For a realistic star, its final catastrophic collapse takes
place in matter of seconds. A star that lived millions of
years thus collapses in only tens of seconds. In the very last
fraction of a microsecond, almost a quarter of its total mass
must be emitted due to quantum effects, and therefore this
would appear like a massive, abrupt burst to an external observer
far away. Typically, such a burst will also carry with it specific
signatures of quantum effects taking place in such ultra-dense
regions. In our case, these included a sudden dip in the
intensity of emission just before the final burstlike evaporation
due to quantum gravity.

The question is, whether such unique astrophysical signatures
can be detected by modern experiments, and if so, what they tell
on quantum gravity, and if there are any new insights into
other aspects of cosmology and fundamental theories such as
string theory.

The key point is, because the very final ultra-dense regions
of the star are no longer hidden within a horizon as in the black
hole case, the exciting possibility of observing these quantum
effects arises now, independently of the quantum gravity theory
used. An astrophysical connection to extreme high energy
phenomena in the universe, such as the gamma-rays bursts that defy
any explanations so far, may not be ruled out.

Such a resolution of naked singularity through quantum gravity
could be a solution to some of the paradoxes mentioned above. 
Then, whenever
a massive star undergoes a gravitational collapse, this might
create a laboratory for quantum gravity in the form of a
{\it Quantum Star}
(see e.g. Joshi, 2009),
that we may be able to possibly access.
This would also suggest intriguing connections to high
energy astrophysical phenomena. The present situation poses
one of the most interesting
challenges which have emerged through the recent work on
gravitational collapse.

We hope the considerations here have shown that gravitational
collapse, which essentially is the investigation of dynamical
evolutions of matter fields
under the force of gravity in the spacetime,
provides one of the most exciting research frontiers in gravitation
physics and high energy astrophysics.
In our view, there is a scope therefore for both theoretical
as well as numerical investigations in these frontier areas, which
may have much to tell for our quest on basic issues in quantum
gravity, fundamental physics and gravity theories, and towards
the expanding frontiers of modern high energy astrophysical
observations.
\bigskip

{\bf References}

Hawking S.W. and Ellis G. F. R., {\it The Large scale structure of 
spacetime}, CUP, Cambridge (1973). 

Joshi P. S., {\it Global aspects in gravitation and cosmology}, 
Clarendon Press, (1993).

Joshi P. S., {\it Gravitational collapse and spacetime singularities},
CUP, Cambridge (2008).

Joshi P. S. {\it Naked Singularities}, Scientific American {\bf 300},
36 (2009).

Joshi P. S. and Dwivedi I. H., Phys. Rev.D 47, 5357, (1993).

Joshi P. S. and Malafarina D. Int.J.Mod.Phys. {\bf D20}, 2641 (2011).

Joshi, P. S., Malafarina D. and Ramesh Narayan, Class. Quantum
Grav. {\bf 28}, 235018 (2011).

Ori A. and Piran T., Phys. Rev. Lett. {\bf 59}, 2137 (1987).

Patil M. and Joshi P. S., Class. and Quant. Grav. {\bf 28}, 
235012 (2011).

Pugliese D., Quevedo H. and Ruffini R., arXiv:1105.2959 [gr-qc] (2011).

Raychaudhuri, A. K., Phys. Rev. {\bf 98}, 1123 (1955).

Thorne K. S. in {\it Magic without Magic--John archibald Wheeler},(ed.
J. Clauder), W. H. Freeman, New York (1972).

\end{document}